\documentclass[sn-chicago]{sn-jnl}% Chicago-based Humanities Reference Style
\jyear{2021}%

\theoremstyle{thmstyleone}%
\newtheorem{theorem}{Theorem}%  meant for continuous numbers
\newtheorem{lemma}{Lemma}%  meant for continuous numbers
\theoremstyle{thmstyletwo}%
\newtheorem{corollary}{Corollary}%

\theoremstyle{thmstylethree}%

\newcommand{\orcid}[1]{\href{https://orcid.org/#1}{\includegraphics[width=12pt]{ORCIDiD_icon24x24.png}}}

\begin{document}

\title{Quantum epistemology and constructivism}

\author[1]{\small \fnm{Patrick} \sur{Fraser}}%\email{p.fraser@mail.utoronto.ca}

\author[2]{\small \fnm{Nuriya} \sur{Nurgalieva}}

\author[2]{\small \fnm{L{\'i}dia} \sur{del Rio}}

\affil[1]{\small \textit{\orgdiv{Department of Philosophy}, \orgname{University of Toronto}, \orgaddress{\city{Toronto}, \country{Canada}}}}

\affil[2]{\small \textit{\orgdiv{Institute for Theoretical Physics}, \orgname{ETH Zurich}, \orgaddress{\city{Zurich}, \country{Switzerland}}}}

\affil[]{\small Corresponding Author: p.fraser@mail.utoronto.ca}

%\author*[1]{\fnm{Patrick} \sur{Fraser}}\email{p.fraser@mail.utoronto.ca}

%\author[2]{\fnm{Nuriya} \sur{Nurgalieva}}

%\author[2]{\fnm{L{\'i}dia} \sur{del Rio}}

%\affil[1]{\orgdiv{Department of Philosophy}, \orgname{University of Toronto}, \orgaddress{\city{Toronto}, \country{Canada}}}

%\affil[2]{\orgdiv{Institute for Theoretical Physics}, \orgname{ETH Zurich}, \orgaddress{\city{Zurich}, \country{Switzerland}}}


\abstract{Constructivist epistemology posits that all truths are knowable. One might ask to what extent constructivism is compatible with naturalized epistemology and knowledge obtained from inference-making using successful scientific theories. If quantum theory correctly describes the structure of the physical world, and if quantum theoretic inferences about which measurement outcomes will be observed with unit probability count as knowledge, we demonstrate that constructivism cannot be upheld. Our derivation is compatible with both intuitionistic and quantum propositional logic. This result is implied by the Frauchiger-Renner theorem, though it is of independent importance as well.}




\maketitle

\section{Introduction}

There are many different perspectives available in contemporary epistemology, and it is not obvious which primitive epistemological commitments are mutually compatible. In this paper, we compare two epistemological perspectives---constructivism and naturalism---and show them to be mutually inconsistent in Section~\ref{sec:our-result} by appealing to the epistemic content of quantum theory on the naturalistic view. We then discuss the significance of this result and show it blocks the derivation of Fitch's paradox within the naturalistic (quantum) epistemic setting in Section~\ref{sec:discussion}.

\subsection{Constructivist epistemology}
Constructivism is the view that truth is grounded in knowability: for a proposition to be true, it must be concretely or constructively demonstrable, and through the contingent possibility that this demonstration could be witnessed by some epistemic agent, knowable (see~\cite{dummett_fitchs_2009,bridges_constructive_2018}). This view is generally expressed as a position about the meaning of mathematical terms and the allowable warrants of mathematical inference (whence it is closely related to intuitionistic logic, cf.~\cite{dummett_elements_2000}). For instance, the constructivist would hold that `there exists an $x$ such that $F(x)$' is true just in case there is some explicit algorithm that could find an $x$ with property $F$ (whence an epistemic agent could, in principle, verify that such an $x$ exists and hence know it). However, the constructivist position may generalize to a hypothesis about the relation between truth, knowledge, and knowability \textit{simpliciter}. It is this general constructivist epistemology that we here bring under scrutiny.

A standard approach for studying a theory of epistemology is to \textit{axiomatize} knowledge as a kind of (modal) logical operator with particular properties determined by that theory and to then investigate their consequences or interpret them via some semantics (see~\cite{hintikka_knowledge_1962,stalnaker_logics_2006,van_ditmarsch_dynamic_2007,dummett_fitchs_2009,van_ditmarsch_handbook_2015,baltag_dynamic_2016,baltag_topological_2019,bjorndahl_logic_2020}). In the present context, we are concerned with the relation between truth, knowledge, and knowability. Thus, we consider a multi-modal propositional language $\mathcal{L}=\{\{K_\alpha\}_{\alpha\in A},\Box,\Diamond,\neg,\to,\land,\lor\}$ where $\Box$, $\Diamond$, and $\{K_\alpha\}$ are all modal operators and each $\alpha$ designates an agent (in some specified collection $A$). We do not yet axiomatize this language, nor do we require that its non-modal formulas resemble \textit{classical} propositional logic (keeping open the possibility of having an underlying \textit{quantum} propositional logic). We understand formulas of the form $K_\alpha\phi$, $\Diamond\phi$, and $\Diamond K_\alpha\phi$ to mean ``$\alpha$ knows that $\phi$,'' ``it is possible that $\phi$,'' and ``it is knowable to $\alpha$ that $\phi$,'' respectively. This language is expressive enough to accommodate knowledge and knowability, and so is sufficient for our analysis of constructivist epistemology.

Consider now the following axiom schemas, where $\phi$ ranges over all well-formed $\mathcal{L}$-formulas and $\alpha$ ranges over $A$:

\begin{itemize}
    \item \textbf{CONST} $\phi\to\Diamond K_\alpha\phi$. Every truth is knowable. This expresses a constructivist view that all truths are concretely demonstrable. %(see~\cite{dummett_fitchs_2009})
    \item \textbf{KCONT} $\neg\Diamond K_\alpha(\phi\land \neg K_\alpha\phi)$. In expressing what they know, no agent can ever assert ``$\phi$ but I do not know that $\phi$.'' This blocks the epistemic version of Moore's paradox by requiring internal knowledge continuity (see~\cite{moore_moores_1993,hintikka_knowledge_1962}).
    \item \textbf{DIST} $(K_\alpha\phi\land K_\alpha\psi)\to K_\alpha(\phi\land\psi)$. If an agent knows two things separately, then she also knows their conjunction.
    \item \textbf{NCK} $\neg K_\alpha(\phi\land\neg\phi)$. No agent has knowledge of a contradiction.
\end{itemize}

Let $\Gamma_{\text{Const}}$ denote the set of all instances of the schemas \textbf{CONST}, \textbf{KCONT}, and let $\Gamma_{\text{Const}}^+$ denote $\Gamma_{\text{Const}}$ together with the set of all instances of the schemas \textbf{DIST} and \textbf{NCK} (so $\Gamma_{\text{Const}}\subseteq\Gamma_{\text{Const}}^+$). Then we shall call the theory of epistemology that is adequately accounted for by $\Gamma_{\text{Const}}$ \textit{constructivism}, and by adding the comparatively weak assumptions \textbf{DIST} and \textbf{NCK}, we shall call the theory of epistemology accounted for by $\Gamma_{\text{Const}}^+$ \textit{enriched constructivism}.

\subsection{Naturalistic epistemology}
Another view that need not be immediately incompatible with constructivism is a weak form of \textit{naturalistic epistemology}. While naturalistic epistemology comes in many forms, we shall define this position as the following assertion: inferences made with certainty about the natural world using the laws governing our best, empirically adequate scientific theories constitute knowledge claims that are possible for agents to know by reasoning about the world only using those theories. We shall adopt this perspective here.

As a general epistemological principle, this naturalistic view may be refined in many ways depending on what scientific theories one has available. We shall be particularly interested in \textit{complete} physical theories. Following~\cite{einstein_can_1935}, a complete physical theory is one whose terms refer to all the actually existing ontological features---entities, properties, relations, etc.---of the physical world, and whose laws determine all true facts about those features. If a theory of epistemology is adequate for describing the knowledge of agents who reason about the world only using inferences warranted by a theory $T$ (whom we shall label $A_T$), then if this theory of epistemology is to be consistent with constructivism, $T$ must be complete. For if $T$ is \textit{not} complete, then there are true facts about the physical world which cannot be described by $T$, and so cannot possibly be known by agents in $A_T$, violating \textbf{CONST}. This should be intuitive: a complete theory, if empirically adequate, is apt for being a fundamental theory, and so the inferences it makes with certainty are reasonably thought to be justifiably true. Completeness is not, however, sufficient to guarantee that naturalistic epistemology is consistent with constructivism, for it need not entail that all true facts about the actual world (all of which may be \textit{described} by $T$) can be \textit{inferred} by $A_T$ agents, and so there may still exist truths expressible in $T$ that are unknowable to $A_T$ agents.

\subsection{Quantum theory}

We now turn to quantum theory $T_{\text{QM}}$ which, for our present purposes, is merely a kind of inferential calculus found in physics textbooks (e.g.~\cite{von_neumann_mathematical_1955,nielsen_quantum_2010}) which, under the right application, can be used to represent material systems (that is, we leave open the interpretation of quantum theory). Quantum theory is the best, most empirically adequate physical theory available to us.\footnote{This is made clear by the far-reaching and highly successful predictions of the Standard Model of particle physics, condensed matter physics, nanoscience, and quantum chemistry, among other things.} Thus, on the naturalistic perspective adopted here, inferences made by agents in $A_{\text{QM}}$ (henceforth quantum agents)---such as inferences about definite measurement outcomes---constitute knowledge. Let us therefore label as `quantum epistemology' any descriptively adequate theory of knowledge that can consistently describe the knowledge possessed by quantum agents about quantum measurements.

Such a theory of quantum epistemology would be a species of naturalistic epistemology, and so would only be compatible with constructivism if quantum theory is complete. The (in)completeness of quantum theory has been the subject of significant historical debate (see~\cite{einstein_can_1935,bohr_can_1935,bohm_suggested_1952,everett_relative_1957,bell_einstein_1964,spekkens_evidence_2007,colbeck_no_2011,pusey_reality_2012,wallace_emergent_2012,leifer_is_2014,mazurek_experimentally_2017}), but if quantum theory is incomplete, quantum epistemology is trivially inconsistent with constructivism. Thus, suppose for the sake of argument that quantum theory is complete. Is quantum epistemology then consistent with (enriched) constructivism? We show it is not.

\section{An Argument From Quantum Theory}\label{sec:our-result}

\subsection{The Frauchiger-Renner theorem}
Suppose that the fundamental dynamics of quantum theory are unitary (e.g. given by the Schr\"odinger equation) and that many quantum systems may be decomposed into subsystems.\footnote{Restricting the discussion to this general case addresses the concerns raised in~\cite{nurgalieva_inadequacy_2019} and~\cite{lazarovici_how_2019}.} A recent result then asserts the following:

\begin{theorem}[\cite*{frauchiger_quantum_2018}]\label{thm:FR}
Not all of the following are true:
\begin{itemize}
    \item \textbf{Q} If the quantum-theoretic Born rule predicts that a measurement outcome will obtain with probability one, it is possible for a quantum agent to know that this measurement outcome will obtain.
    \item \textbf{S} If a quantum agent knows that a particular measurement outcome will obtain, then they cannot also know that a different measurement outcome will obtain.
    \item \textbf{C} Whenever a quantum agent $\alpha$ knows that another quantum agent $\beta$ knows that some measurement outcome will obtain, $\alpha$ also knows that this measurement outcome will obtain.
\end{itemize}
\end{theorem}

This result has led to significant controversy (see~\cite{healey_quantum_2018,lazarovici_how_2019,sudbery_hidden_2019,vilasini_multi-agent_2019,kastner_unitary-only_2020,nurgalieva_testing_2020,waaijer_relational_2021}). However, whereas there is disagreement about the \textit{interpretation} of this result and disagreement about which assumption fails, there is a general consensus that the derivation of the result is itself correct.

One can use the epistemic language $\mathcal{L}$ to describe the knowledge claims about propositions that express statements about measurement outcomes made by quantum agents. Here, the underlying propositional logic is \textit{quantum} propositional logic as described by~\cite{birkhoff_logic_1936}. We may then re-express Theorem~\ref{thm:FR} in $\mathcal{L}$ as follows: Let $P_{\text{QM}}$ be the set of all propositions about quantum measurement outcomes and say that an agent knows that $p\in P_{\text{QM}}$ just in case, given her existing knowledge, she can infer that $p$ with probability 1 using quantum theory. Quantum knowledge is then such that for some agents $\alpha,\beta\in A_{\text{QM}}$, there is a quantum proposition $p\in P_{\text{QM}}$ for which at least one of the following axioms is false:

\begin{equation*}
    \textbf{EQ}\quad p\to\Diamond K_\alpha p,\qquad \textbf{ES}\quad \neg(K_\alpha p\land K_\alpha(\neg p)),\qquad \textbf{EC}\quad \neg(K_\alpha K_\beta p\land\neg K_\alpha p).
\end{equation*}

Note that the axiom schemas \textbf{EQ}, \textbf{ES}, and \textbf{EC} are \textit{not} defined inductively over $\mathcal{L}$-formulas; they are only defined for arbitrary atomic proposition $p$.

\subsection{Our result}
Given this, our novel result (proven in Appendix~\ref{app:proofs}) is as follows:

\begin{theorem}\label{thm:our-result}
Any counter-instance to one of \textbf{EQ}, \textbf{ES}, or \textbf{EC} is inconsistent with $\Gamma_{\text{Const}}^+$. In particular, a counter-instance of \textbf{EQ} is inconsistent with \textbf{CONST}, a counter instance of \textbf{ES} is inconsistent with $\Gamma^+_{\text{Const}}\setminus\Gamma_{\text{Const}}$, and a counter instance of \textbf{EC} is inconsistent with $\Gamma^+_{\text{Const}}$.
\end{theorem}

The derivation of this result is intuitionistically valid and so satisfies the standards for proof required by the constructivist. Since quantum epistemology must satisfy the Frauchiger-Renner theorem, we therefore conclude that quantum epistemology is inconsistent with enriched constructivism. We also have the following:

\begin{corollary}\label{cor:Q-or-C-false}
If quantum epistemology satisfies \textbf{DIST} and \textbf{NCK}, then it is inconsistent with $\Gamma_{\text{Const}}$ and one of \textbf{Q} or \textbf{C} is false of quantum theory.
\end{corollary}

\section{Discussion}\label{sec:discussion}
Theorem~\ref{thm:our-result} shows that enriched constructivism is incompatible with quantum epistemology and puts limits on the viable resolutions to the Frauchiger-Renner paradox, given one's epistemic commitments via Corollary~\ref{cor:Q-or-C-false}.

From Corollary~\ref{cor:Q-or-C-false}, if quantum epistemology has minimal assumed epistemic structure, then either \textbf{Q} or \textbf{C} must fail,\footnote{One might argue that on a many-worlds interpretation of quantum mechanics (cf.~\cite{everett_relative_1957,wallace_emergent_2012}), the \textbf{S} assumption is the one violated since all measurement outcomes are observed in some branch. However, most many-worlds interpretations distinguish agents whose experiences lie within different branches and so even on a many-worlds picture, \textbf{S} is upheld since all agents see only one outcome; \textbf{C} is more directly challenged.} and otherwise, one must deny either \textbf{DIST} or \textbf{NCK}. Since \textbf{Q} is essentially the assumption that quantum theoretic predictions made with certainty constitute knowledge, if one maintains that quantum theory is the best, most empirically adequate theory available, \textbf{Q} must be upheld. Thus, it seems most natural to reject \textbf{C}. Our result therefore provides a principled, interpretation-independent argument for which assumption must be given up. Importantly, \textbf{C} resembles a form of the principle that has elsewhere been called \textit{observer independence of facts} or \textit{local friendliness} which have already been challenged by other research for other reasons (see~\cite{brukner_no-go_2018,bong_strong_2020}).

Our result is also relevant to broader issues in epistemic logic: from $\Gamma_{\text{Const}}$, one can derive the knowability paradox due to~\cite{fitch_logical_2009}, according to which if all truths are knowable, then all truths are known. Taking $\phi$ to range over all $\mathcal{L}$-formulas and $\alpha$ to range over all agents in $A$, Fitch's paradox is:

\begin{theorem}[\cite{dummett_fitchs_2009}]\label{thm:fitch}
$\Gamma_{\text{Const}}\vdash\phi\to K_\alpha\phi$.
\end{theorem}

This theorem is not intuitionistically valid (and so unsatisfying for the constructivist), however, the theorem $\Gamma_{\text{Const}}\vdash\phi\to\neg\neg K_\alpha\phi$ is intuitionistically valid (see Appendix~\ref{app:proofs}). Given that actual knowledge seems to be contingent in a way that truth is not, this result seems wrong (especially since if knowledge is factive, knowledge and truth would then become definitionally equivalent). Traditional epistemology has acknowledged the difficulties of conceding Fitch's paradox (see~\cite{williamson_intutionism_1982,edgington_paradox_1985,williamson_paradox_1987,percival_fitch_1990,kvanvig_knowability_1995,tennant_taming_1997,hand_tennant_1999,dividi_knowability_2001,fitch_logical_2009,dummett_fitchs_2009,restall_not_2009,edgington_possible_2010}). From our result, we find that if \textbf{DIST} and \textbf{NCK} are true of quantum epistemology, then quantum epistemology does not admit Fitch's paradox (at least not via its usual deduction). Thus, we provide a positive, naturalistic remedy to a classical epistemological challenge.

\backmatter

\bmhead{Acknowledgments}

PF is supported in part by funding from the Social Sciences and Humanities Research Council.
NN and LdR acknowledge support from the Swiss National Science Foundation through SNSF project No.\ $200021\_188541$ and through the the National Centre of Competence in Research \emph{Quantum Science and Technology} (QSIT). LdR further acknowledges support from the FQXi grant \emph{Consciousness in the Physical World}.
LdR is grateful for the hospitality of Perimeter Institute where part of this work was carried out. Research at Perimeter Institute is supported in part by the Government of Canada through the Department of Innovation, Science and Economic Development and by the Province of Ontario through the Ministry of Colleges and Universities. %Discussions for this paper started at the summer school \href{http://foundations.ethz.ch}{Solstice of Foundations 2019}, supported by the Pauli Center for Theoretical Studies in Zurich.
This project was made possible by a chance encounter between PF and LdR at Caf\'e Pamenar in Toronto. We thank Niels Linnemann, Thomas de Saegher, Wayne Myrvold, and Bas van Fraassen, as well as the audiences at the \textit{Foundations of Physics} meeting in Paris and the \textit{CQIQC} seminar in Toronto for valuable comments.

\begin{appendices}

\section{Proofs}
\label{app:proofs}

We do not assume that $\mathcal{L}$ satisfies the axioms of classical propositional logic, for we want it also to be compatible with an underlying quantum propositional logic while only allowing for intuitionistically valid deductions. However, we do stipulate that it satisfies the following axiom schemas (whose instances shall collectively be denoted $\Sigma$) for all $\mathcal{L}$-formulas $\phi$ and $\psi$:

\begin{alignat*}{3}
    \text{Double Negation Introduction (\textbf{DNI}):}\quad&&&\vdash\phi\to\neg\neg\phi\\
    \text{Triple Negation Elimination (\textbf{TNE}):}\quad&&&\vdash\neg\neg\neg\phi\to\neg\phi\\
    \text{Conjunction Negation Distribution (\textbf{CND}):}\quad&&&\vdash(\neg\neg\phi\land\neg\neg\psi)\to\neg\neg(\phi\land\psi)\\
    \text{Conjunction Introduction (\textbf{CI}):}\quad&\{\phi,\psi\}&&\vdash\phi\land\psi\\
    \text{Conjunction Elimination (\textbf{CE}):}\quad&&&\vdash\phi\land\psi\to\phi,\quad\vdash\phi\land\psi\to\psi\\
    \text{Contraposition:}\quad&&&\vdash(\phi\to\psi)\to(\neg\psi\to\neg\phi)\\
    \text{Propositional Identity:}\quad&&&\vdash\neg(\phi\land\neg\psi)\to(\phi\to\neg\neg\psi)
\end{alignat*}

These axioms are quite minimal (indeed, we do not posit any modal axioms whatsoever) and insufficient on their own to prove completeness. This is a positive feature of the generality of our analysis. Note that all of these axioms are intuitionistically valid and are valid in quantum propositional logic as well (so they can be used in the proof of our main result while being consistent with a constructivist account of logic). The only rule of inference we shall assume is \textit{modus ponens} (we do not assume the necessitation rule).

Before proving our main theorem, we prove Theorem~\ref{thm:fitch} (which entails Fitch's paradox) and several lemmas. Defining $\vdash$ with respect to $\Sigma$, Theorem~\ref{thm:fitch} asserts that for any $\mathcal{L}$-formula $\phi$ and agent $\alpha\in A$, $\Gamma_{\text{Const}}\vdash\phi\to\neg\neg K_\alpha\phi$:

\begin{proof}
For any $\mathcal{L}$-formula $\phi$ we have:

\begin{alignat*}{2}
        1.\quad&(\phi\land\neg K_\alpha\phi)\to\Diamond K_\alpha(\phi\land\neg K_\alpha\phi)\quad&\text{\textbf{CONST}.}\\
        2.\quad&\neg\Diamond K_\alpha(\phi\land\neg K_\alpha\phi)\to\neg(\phi\land\neg K_\alpha\phi)\quad&\text{Contraposition.}\\
        3.\quad&\neg\Diamond K_\alpha(\phi\land \neg K_\alpha\phi)\quad\qquad&\text{\textbf{KCONT}.}\\
        4.\quad&\neg(\phi\land\neg K_\alpha\phi)\quad&\text{2, 3, Modus Ponens.}\\
        5.\quad&\neg(\phi\land\neg K_\alpha\phi)\to(\phi\to\neg\neg K_\alpha\phi)\quad&\text{Propositional Identity.}\\
        6.\quad&\phi\to\neg\neg K_\alpha\phi\quad&\text{4, 5, Modus Ponens.}
\end{alignat*}
\end{proof}

We shall denote by \textbf{FITCH} the set of all instances of $\phi\to\neg\neg K_\alpha\phi$ for all $\mathcal{L}$-formulas $\phi$ and agents $\alpha\in A$. We now prove several lemmas, taking $\phi$ and $\psi$ to range over all $\mathcal{L}$-formulas, and $\alpha$ and $\beta$ to range over all agents in $A$.

\begin{lemma}\label{lem:neg-phi}
$\Gamma_{\text{Const}}\cup\{\neg K_\alpha\phi\}\vdash\neg\phi$.
\end{lemma}

\begin{proof}
\begin{alignat*}{2}
    1.\quad&\neg K_\alpha\phi\quad&\text{Assumption.}\\
    2.\quad&\phi\to\neg\neg K_\alpha\phi\quad&\text{\textbf{FITCH} for $\alpha$.}\\
    3.\quad&\neg\neg\neg K_\alpha\phi\to\neg\phi\quad&\text{2, Contraposition.}\\
    4.\quad&\neg K_\alpha\phi\to\neg\neg\neg K_\alpha\phi\quad&\text{\textbf{DNI}.}\\
    5.\quad&\neg\neg\neg K_\alpha\phi\quad&\text{1, 4, Modus Ponens.}\\
    6.\quad&\neg\phi\quad&\text{3, 5, Modus Ponens.}\\
\end{alignat*}
\end{proof}

\begin{lemma}\label{lem:K-KK}
$\Gamma_{\text{Const}}\cup\{\neg\neg K_\beta\phi\}\vdash \neg\neg K_\alpha K_\beta\phi$.
\end{lemma}

\begin{proof}
\begin{alignat*}{2}
    1.\quad&\neg\neg K_\beta\phi\quad&\text{Assumption.}\\
    2.\quad& K_\beta\phi\to \neg\neg K_\alpha K_\beta\phi&\text{\textbf{FITCH} for $\alpha$.}\\
    3.\quad& (K_\beta\phi\to \neg\neg K_\alpha K_\beta\phi)\to(\neg\neg\neg K_\alpha K_\beta\phi\to\neg K_\beta\phi)&\text{Contraposition.}\\
    4.\quad& \neg\neg\neg K_\alpha K_\beta\phi\to\neg K_\beta\phi&\text{2, 3, Modus Ponens.}\\
    5.\quad& (\neg\neg\neg K_\alpha K_\beta\phi\to\neg K_\beta\phi)\to(\neg\neg K_\beta\phi\to\neg\neg\neg\neg K_\alpha K_\beta\phi)&\text{Contraposition.}\\
    6.\quad& \neg\neg K_\beta\phi\to\neg\neg\neg\neg K_\alpha K_\beta\phi&\text{4, 5, Modus Ponens.}\\
    7.\quad&\neg\neg\neg\neg K_\alpha K_\beta\phi&\text{1, 6, Modus Ponens.}\\
    8.\quad&\neg\neg\neg\neg K_\alpha K_\beta\phi\to\neg\neg K_\alpha K_\beta\phi&\text{\textbf{TNE}.}\\
    9.\quad&\neg\neg K_\alpha K_\beta\phi&\text{7, 8, Modus Ponens.}\\
\end{alignat*}
\end{proof}

\begin{lemma}\label{lem:negneg-dist}
$\Gamma_{\text{Const}}\cup\{(\neg\neg K_\alpha \phi)\land(\neg\neg K_\alpha \psi)\}\vdash\neg\neg K_\alpha(\phi\land\psi)$.
\end{lemma}

\begin{proof}
\begin{alignat*}{2}
    1.\quad& (\neg\neg K_\alpha \phi)\land(\neg\neg K_\alpha \psi)&\text{Assumption.}\\
    2.\quad& ((\neg\neg K_\alpha \phi)\land(\neg\neg K_\alpha \psi))\to\neg\neg(K_\alpha \phi\land K_\alpha\psi)&\text{\textbf{CND}.}\\
    3.\quad& \neg\neg(K_\alpha \phi\land K_\alpha\psi)&\text{1, 2, Modus Ponens.}\\
    4.\quad& (K_\alpha\phi\land K_\alpha\psi)\to K_\alpha(\phi\land\psi)&\text{\textbf{DIST}.}\\
    5.\quad& ((K_\alpha\phi\land K_\alpha\psi)\to K_\alpha(\phi\land\psi))&\text{Contraposition.}\\
           & \qquad\qquad \to( \neg K_\alpha(\phi\land\psi)\to\neg(K_\alpha\phi\land K_\alpha\psi))&\\
    6.\quad& \neg K_\alpha(\phi\land\psi)\to\neg(K_\alpha\phi\land K_\alpha\psi)&\text{4, 5, Modus Ponens.}\\
    7.\quad& (\neg K_\alpha(\phi\land\psi)\to\neg(K_\alpha\phi\land K_\alpha\psi))&\text{Contraposition.}\\
           & \qquad\qquad\to(\neg\neg(K_\alpha\phi\land K_\alpha\psi)\to\neg\neg K_\alpha(\phi\land\psi))&\\
    8.\quad& \neg\neg(K_\alpha\phi\land K_\alpha\psi)\to\neg\neg K_\alpha(\phi\land\psi)&\text{6, 7, Modus Ponens.}\\
    9.\quad&\neg\neg K_\alpha(\phi\land\psi)&\text{3, 8, Modus Ponens.}
\end{alignat*}
\end{proof}

The proof of our main result, Theorem~\ref{thm:our-result}, is then as follows.

\begin{proof}
We must prove three separate results. First, a counter-instance of \textbf{EQ} is inconsistent with \textbf{CONST}. If we assume that \textbf{EQ} is violated, then there is some agent $\alpha\in A_{\text{QM}}$ and some quantum proposition $p\in P_{\text{QM}}$ such that $\neg(p\to\Diamond K_\alpha p)$ (a formula which which we shall denote by $\neg\textbf{EQ}$). Letting $\phi_p:=p\to\Diamond K_\alpha p$ we have:

\begin{alignat*}{2}
    1.\quad& p\to\Diamond K_\alpha p &\text{\textbf{CONST}.}\\
    2.\quad& \neg(p\to\Diamond K_\alpha p) &\neg\text{\textbf{EQ}.}\\
    3.\quad& \phi_p\land\neg\phi_p &\text{1, 2, \textbf{CI}.}\\
\end{alignat*}

\noindent Thus, $\textbf{CONST}\cup\{\neg\textbf{EQ}\}\vdash\perp$.

Next, a counter instance of \textbf{ES} is inconsistent with $\Gamma^+_{\text{Const}}\setminus\Gamma_{\text{Const}}$. If we assume that \textbf{ES} is violated, then there is some agent $\alpha\in A_{\text{QM}}$ and some quantum proposition $p\in P_{\text{QM}}$ such that $\neg\neg(K_\alpha p\land K_\alpha(\neg p))$ (a formula which we shall denote by $\neg\textbf{ES}$). Letting $\psi_p:=\neg K_\alpha(p\land \neg p)$, we have:

\begin{alignat*}{2}
    1.\quad& \neg K_\alpha(p\land \neg p) &\text{\textbf{NCK}.}\\
    2.\quad& \neg\neg(K_\alpha p\land K_\alpha(\neg p)) &\neg\text{\textbf{ES}.}\\
    3.\quad& \neg\neg K_\alpha(p\land \neg p) &\text{2, \textbf{DIST}.}\\
    4.\quad& \psi_p\land\neg\psi_p &\text{1, 3, \textbf{CI}.}\\
\end{alignat*}

\noindent Thus, $\textbf{NCK}\cup\textbf{DIST}\cup\{\neg\textbf{ES}\}\vdash\perp$ (where $\Gamma^+_{\text{Const}}\setminus\Gamma_{\text{Const}}=\textbf{NCK}\cup\textbf{DIST}$).

Finally, a counter instance of \textbf{EC} is inconsistent with $\Gamma^+_{\text{Const}}$. If we assume that \textbf{EC} is violated, then there are some agents $\alpha,\beta\in A_{\text{QM}}$ and some quantum proposition $p\in P_{\text{QM}}$ such that $\neg\neg(K_\alpha K_\beta p\land\neg K_\alpha p)$ (a formula we shall denote by $\neg$\textbf{EC}). Letting $\xi_p:=\neg K_\alpha((K_\beta p\land K_\beta(\neg p))\land\neg(K_\beta p\land K_\beta(\neg p)))$, we have:

\begin{alignat*}{2}
    1.\quad& \neg\neg(K_\alpha K_\beta p\land\neg K_\alpha p)&\text{$\neg$\textbf{EC}.}\\
    2.\quad& \neg\neg(K_\alpha K_\beta p\land\neg K_\alpha p)\to(\neg\neg K_\alpha K_\beta p\land\neg\neg\neg K_\alpha p)&\text{\textbf{CND}.}\\
    3.\quad& \neg\neg K_\alpha K_\beta p\land\neg\neg\neg K_\alpha p&\text{1, 2, Modus Ponens.}\\
    4.\quad& \neg\neg K_\alpha K_\beta p&\text{3, \textbf{CE}.}\\
    5.\quad& \neg\neg\neg K_\alpha p&\text{3, \textbf{CE}.}\\
    6.\quad& \neg\neg\neg K_\alpha p\to\neg K_\alpha p&\text{\textbf{TNE}.}\\
    7.\quad& \neg K_\alpha p&\text{5, 6, Modus Ponens.}\\
    8.\quad& \neg p&\text{7, Lemma~\ref{lem:neg-phi}.}\\
    9.\quad& \neg p\to \neg\neg K_\beta(\neg p)&\text{\textbf{FITCH} for $\beta$.}\\
    10.\quad& \neg\neg K_\beta(\neg p)&\text{8, 9, Modus Ponens.}\\
    11.\quad& \neg\neg K_\alpha K_\beta(\neg p)&\text{10, Lemma~\ref{lem:K-KK}.}\\
    12.\quad& (\neg\neg K_\alpha K_\beta p)\land(\neg\neg K_\alpha K_\beta(\neg p))&\text{4, 11, \textbf{CI}.}\\
    13.\quad& \neg\neg K_\alpha (K_\beta p\land K_\beta(\neg p))&\text{12, Lemma~\ref{lem:negneg-dist}.}\\
    14.\quad& (K_\beta p\land K_\beta(\neg p))\to K_\beta(p\land\neg p)&\text{\textbf{DIST}.}\\
    15.\quad& ((K_\beta p\land K_\beta(\neg p))\to K_\beta(p\land\neg p))&\text{Contraposition.}\\
            &\qquad\qquad\to(\neg K_\beta(p\land\neg p)\to\neg (K_\beta p\land K_\beta(\neg p)))&\\
    16.\quad& \neg K_\beta(p\land\neg p)\to\neg (K_\beta p\land K_\beta(\neg p))&\text{14, 15, Modus Ponens.}\\
    17.\quad& \neg K_\beta(p\land\neg p)&\text{\textbf{NCK}.}\\
    18.\quad& \neg(K_\beta p\land K_\beta(\neg p))&\text{16, 17, Modus Ponens.}\\
    19.\quad& \neg(K_\beta p\land K_\beta(\neg p))\to\neg\neg K_\alpha(\neg(K_\beta p\land K_\beta(\neg p)))&\text{\textbf{FITCH} for $\alpha$.}\\
    20.\quad& \neg\neg K_\alpha(\neg(K_\beta p\land K_\beta(\neg p)))&\text{18, 19, Modus Ponens.}\\
    21.\quad& (\neg\neg K_\alpha(K_\beta p\land K_\beta(\neg p)))\land(\neg\neg K_\alpha(\neg(K_\beta p\land K_\beta(\neg p))))&\text{13, 20, \textbf{CI}.}\\
    22.\quad& \neg\neg K_\alpha((K_\beta p\land K_\beta(\neg p))\land\neg(K_\beta p\land K_\beta(\neg p)))&\text{21, Lemma~\ref{lem:negneg-dist}.}\\
    23.\quad& \neg K_\alpha((K_\beta p\land K_\beta(\neg p))\land\neg(K_\beta p\land K_\beta(\neg p)))&\text{\textbf{NCK}.}\\
    24.\quad&\xi_p\land\neg\xi_p&\text{22, 23, \textbf{CI}.}
\end{alignat*}

\noindent Thus, $\Gamma^+_{\text{Const}}\cup\{\neg\textbf{EC}\}\vdash\perp$.

\end{proof}

\end{appendices}


\bibliography{FR-Fitch-2}% common bib file

\begin{thebibliography}{}
\providecommand{\doi}[1]{\url{https://doi.org/#1}}
\bibcommenthead

\bibitem[\protect\citeauthoryear{Baltag, Bezhanishvili, Özgün, and
  Smets}{Baltag et~al.}{2019}]{baltag_topological_2019}
Baltag, A., N.~Bezhanishvili, A.~Özgün, and S.~Smets. 2019, April.
\newblock A {Topological} {Approach} to {Full} {Belief}.
\newblock {\em Journal of Philosophical Logic\/}~{\em 48\/}(2): 205--244.
\newblock \doi{10.1007/s10992-018-9463-4} .

\bibitem[\protect\citeauthoryear{Baltag and Renne}{Baltag and
  Renne}{2016}]{baltag_dynamic_2016}
Baltag, A. and B.~Renne. 2016.
\newblock Dynamic {Epistemic} {Logic}, In {\em The {Stanford} {Encyclopedia} of
  {Philosophy}\/} (Winter 2016 ed.).,  ed. Zalta, E.N. Metaphysics Research
  Lab, Stanford University.

\bibitem[\protect\citeauthoryear{Bell}{Bell}{1964}]{bell_einstein_1964}
Bell, J.S. 1964.
\newblock On the {Einstein} {Podolsky} {Rosen} paradox.
\newblock {\em Physics Physique Fizika\/}~{\em 1\/}(3): 195--200.
\newblock \doi{10.1103/PhysicsPhysiqueFizika.1.195} .

\bibitem[\protect\citeauthoryear{Birkhoff and Von~Neumann}{Birkhoff and
  Von~Neumann}{1936}]{birkhoff_logic_1936}
Birkhoff, G. and J.~Von~Neumann. 1936.
\newblock The logic of quantum mechanics.
\newblock {\em Annals of mathematics\/}: 823--843.
\newblock \doi{10.2307/1968621} .

\bibitem[\protect\citeauthoryear{Bjorndahl and Özgün}{Bjorndahl and
  Özgün}{2020}]{bjorndahl_logic_2020}
Bjorndahl, A. and A.~Özgün. 2020, December.
\newblock Logic and {Topology} for {Knowledge}, {Knowability}, and {Belief}.
\newblock {\em The Review of Symbolic Logic\/}~{\em 13\/}(4): 748--775.
\newblock \doi{10.1017/S1755020319000509} .

\bibitem[\protect\citeauthoryear{Bohm}{Bohm}{1952}]{bohm_suggested_1952}
Bohm, D. 1952.
\newblock A {Suggested} {Interpretation} of the {Quantum} {Theory} in {Terms}
  of "{Hidden}" {Variables}. {I} and {II}.
\newblock {\em Physical Review\/}~{\em 85\/}(2): 166--179.
\newblock \doi{10.1103/physrev.85.166} .

\bibitem[\protect\citeauthoryear{Bohr}{Bohr}{1935}]{bohr_can_1935}
Bohr, N. 1935.
\newblock Can {Quantum}-{Mechanical} {Description} of {Physical} {Reality} be
  {Considered} {Complete}?
\newblock {\em Physical Review\/}~{\em 48\/}(8): 696.
\newblock \doi{10.1103/PhysRev.48.696} .

\bibitem[\protect\citeauthoryear{Bong, Utreras-Alarcón, Ghafari, Liang,
  Tischler, Cavalcanti, Pryde, and Wiseman}{Bong
  et~al.}{2020}]{bong_strong_2020}
Bong, K.W., A.~Utreras-Alarcón, F.~Ghafari, Y.C. Liang, N.~Tischler, E.G.
  Cavalcanti, G.J. Pryde, and H.M. Wiseman. 2020, December.
\newblock A strong no-go theorem on the {Wigner}’s friend paradox.
\newblock {\em Nature Physics\/}~{\em 16\/}(12): 1199--1205.
\newblock \doi{10.1038/s41567-020-0990-x} .

\bibitem[\protect\citeauthoryear{Bridges and Palmgren}{Bridges and
  Palmgren}{2018}]{bridges_constructive_2018}
Bridges, D. and E.~Palmgren. 2018.
\newblock Constructive {Mathematics}, In {\em The {Stanford} {Encyclopedia} of
  {Philosophy}\/} (Summer 2018 ed.).,  ed. Zalta, E.N. Metaphysics Research
  Lab, Stanford University.

\bibitem[\protect\citeauthoryear{Brukner}{Brukner}{2018}]{brukner_no-go_2018}
Brukner, C. 2018, April.
\newblock A {No}-{Go} {Theorem} for {Observer}-{Independent} {Facts}.
\newblock {\em Entropy\/}~20.
\newblock \doi{10.3390/e20050350} .

\bibitem[\protect\citeauthoryear{Colbeck and Renner}{Colbeck and
  Renner}{2011}]{colbeck_no_2011}
Colbeck, R. and R.~Renner. 2011, August.
\newblock No extension of quantum theory can have improved predictive power.
\newblock {\em Nature Communications\/}~{\em 2\/}(1): 411.
\newblock \doi{10.1038/ncomms1416} .

\bibitem[\protect\citeauthoryear{DiVidi and Solomon}{DiVidi and
  Solomon}{2001}]{dividi_knowability_2001}
DiVidi, D. and G.~Solomon. 2001.
\newblock Knowability and intuitionistic logic.
\newblock {\em Philosophia\/}~28 .

\bibitem[\protect\citeauthoryear{Dummett}{Dummett}{2000}]{dummett_elements_2000}
Dummett, M. 2000, August.
\newblock {\em Elements of {Intuitionism}\/} (Second ed.).
\newblock Oxford {Logic} {Guides}. Oxford, New York: Oxford University Press.

\bibitem[\protect\citeauthoryear{Dummett}{Dummett}{2009}]{dummett_fitchs_2009}
Dummett, M. 2009.
\newblock Fitch's {Paradox} of {Knowability}, In {\em New essays on the
  knowability paradox},  ed. Salerno, J.,  51--52. Oxford University Press.
\newblock Section: 4.
\newblock \doi{10.1093/acprof:oso/9780199285495.003.0005}.

\bibitem[\protect\citeauthoryear{Edgington}{Edgington}{1985}]{edgington_paradox_1985}
Edgington, D. 1985.
\newblock The {Paradox} of {Knowability}.
\newblock {\em Mind\/}~{\em 94\/}(376): 557--568 .

\bibitem[\protect\citeauthoryear{Edgington}{Edgington}{2010}]{edgington_possible_2010}
Edgington, D. 2010.
\newblock Possible {Knowledge} of {Unknown} {Truth}.
\newblock {\em Synthese\/}~173.
\newblock \doi{10.1007/s11229-009-9675-9} .

\bibitem[\protect\citeauthoryear{Einstein, Podolsky, and Rosen}{Einstein
  et~al.}{1935}]{einstein_can_1935}
Einstein, A., B.~Podolsky, and N.~Rosen. 1935.
\newblock Can {Quantum}-{Mechanical} {Description} of {Physical} {Reality} {Be}
  {Considered} {Complete}?
\newblock {\em Physical Review\/}~{\em 47\/}(10): 777.
\newblock \doi{10.1103/PhysRev.47.777} .

\bibitem[\protect\citeauthoryear{Everett}{Everett}{1957}]{everett_relative_1957}
Everett, H. 1957, July.
\newblock "{Relative} {State}" {Formulation} of {Quantum} {Mechanics}.
\newblock {\em Reviews of Modern Physics\/}~{\em 29\/}(3): 454--462.
\newblock \doi{10.1103/revmodphys.29.454} .

\bibitem[\protect\citeauthoryear{Fitch}{Fitch}{2009}]{fitch_logical_2009}
Fitch, F.B. 2009.
\newblock A {Logical} {Analysis} of {Some} {Value} {Concepts}, In {\em New
  essays on the knowability paradox},  ed. Salerno, J.,  34--41. Oxford
  University Press.
\newblock Section: 2.
\newblock \doi{10.2307/2271594}.

\bibitem[\protect\citeauthoryear{Frauchiger and Renner}{Frauchiger and
  Renner}{2018}]{frauchiger_quantum_2018}
Frauchiger, D. and R.~Renner. 2018.
\newblock Quantum theory cannot consistently describe the use of itself.
\newblock {\em Nature Communications\/}~{\em 9\/}(1): 3711.
\newblock \doi{10.1038/s41467-018-05739-8} .

\bibitem[\protect\citeauthoryear{Hand and Kvanvig}{Hand and
  Kvanvig}{1999}]{hand_tennant_1999}
Hand, M. and J.L. Kvanvig. 1999.
\newblock Tennant on knowability.
\newblock {\em Australasian Journal of Philosophy\/}~{\em 77\/}(4).
\newblock \doi{10.1080/00048409912349191} .

\bibitem[\protect\citeauthoryear{Healey}{Healey}{2018}]{healey_quantum_2018}
Healey, R. 2018.
\newblock Quantum {Theory} and the {Limits} of {Objectivity}.
\newblock {\em Foundations of Physics\/}~48: 1568--1589.
\newblock \doi{10.1007/s10701-018-0216-6} .

\bibitem[\protect\citeauthoryear{Hintikka}{Hintikka}{1962}]{hintikka_knowledge_1962}
Hintikka, J. 1962.
\newblock {\em Knowledge and {Belief}}.
\newblock Ithaca: Cornell University Press.

\bibitem[\protect\citeauthoryear{Kastner}{Kastner}{2020}]{kastner_unitary-only_2020}
Kastner, R.E. 2020.
\newblock Unitary-{Only} {Quantum} {Theory} {Cannot} {Consistently} {Describe}
  the {Use} of {Itself}: {On} the {Frauchiger}–{Renner} {Paradox}.
\newblock {\em Foundations of Physics\/}~50: 441--456.
\newblock \doi{10.1007/s10701-020-00336-6} .

\bibitem[\protect\citeauthoryear{Kvanvig}{Kvanvig}{1995}]{kvanvig_knowability_1995}
Kvanvig, J.L. 1995.
\newblock The {Knowability} {Paradox} and the {Prospects} for {Anti}-{Realism}.
\newblock {\em Noûs\/}~{\em 29\/}(4): 481--500.
\newblock \doi{10.2307/2216283} .

\bibitem[\protect\citeauthoryear{Lazarovici and Hubert}{Lazarovici and
  Hubert}{2019}]{lazarovici_how_2019}
Lazarovici, D. and M.~Hubert. 2019.
\newblock How {Quantum} {Mechanics} can consistently describe the use of
  itself.
\newblock {\em Scientific Reports\/}~{\em 9\/}(470).
\newblock \doi{10.1038/s41598-018-37535-1} .

\bibitem[\protect\citeauthoryear{Leifer}{Leifer}{2014}]{leifer_is_2014}
Leifer, M.S. 2014, November.
\newblock Is the {Quantum} {State} {Real}? {An} {Extended} {Review} of
  {$\psi$}-ontology {Theorems}.
\newblock {\em Quanta\/}~{\em 3\/}(1): 67--155.
\newblock \doi{10.12743/quanta.v3i1.22} .

\bibitem[\protect\citeauthoryear{Mazurek, Pusey, Resch, and Spekkens}{Mazurek
  et~al.}{2017}]{mazurek_experimentally_2017}
Mazurek, M.D., M.F. Pusey, K.J. Resch, and R.W. Spekkens. 2017.
\newblock Experimentally bounding deviations from quantum theory in the
  landscape of generalized probabilistic theories.
\newblock Published: arXiv:1710.05948.

\bibitem[\protect\citeauthoryear{Moore}{Moore}{1993}]{moore_moores_1993}
Moore, G.E. 1993.
\newblock Moore's {Paradox}, In {\em G. {E}. {Moore}: {Selected} {Writings}},
  ed. Baldwin, T. Routledge.
\newblock Section: 207–212.

\bibitem[\protect\citeauthoryear{Nielsen and Chuang}{Nielsen and
  Chuang}{2010}]{nielsen_quantum_2010}
Nielsen, M. and I.~Chuang. 2010.
\newblock {\em Quantum {Computing} and {Quantum} {Information}}.
\newblock Cambridge University Press.

\bibitem[\protect\citeauthoryear{Nurgalieva and {del Rio}}{Nurgalieva and {del
  Rio}}{2019}]{nurgalieva_inadequacy_2019}
Nurgalieva, N. and L.~{del Rio}. 2019.
\newblock Inadequacy of {Modal} {Logic} in {Quantum} {Settings}.
\newblock {\em EPCTS\/}~287: 267--297.
\newblock \doi{10.4204/EPTCS.287.16} .

\bibitem[\protect\citeauthoryear{Nurgalieva and Renner}{Nurgalieva and
  Renner}{2020}]{nurgalieva_testing_2020}
Nurgalieva, N. and R.~Renner. 2020, July.
\newblock Testing quantum theory with thought experiments.
\newblock {\em Contemporary Physics\/}~{\em 61\/}(3): 193--216.
\newblock \doi{10.1080/00107514.2021.1880075} .

\bibitem[\protect\citeauthoryear{Percival}{Percival}{1990}]{percival_fitch_1990}
Percival, P. 1990.
\newblock Fitch and {Intuitionistic} {Knowability}.
\newblock {\em Analysis\/}~{\em 50\/}(3).
\newblock \doi{10.2307/3328541} .

\bibitem[\protect\citeauthoryear{Pusey, Barrett, and Rudolph}{Pusey
  et~al.}{2012}]{pusey_reality_2012}
Pusey, M.F., J.~Barrett, and T.~Rudolph. 2012, June.
\newblock On the reality of the quantum state.
\newblock {\em Nature Physics\/}~{\em 8\/}(6): 475--478.
\newblock \doi{10.1038/nphys2309} .

\bibitem[\protect\citeauthoryear{Restall}{Restall}{2009}]{restall_not_2009}
Restall, G. 2009.
\newblock Not {Every} {Truth} {Can} {Be} {Known} (at least, not all at once),
  In {\em New essays on the knowability paradox},  ed. Salerno, J.,  51--52.
  Oxford University Press.
\newblock Section: 21.
\newblock \doi{10.1093/acprof:oso/9780199285495.003.0022}.

\bibitem[\protect\citeauthoryear{Spekkens}{Spekkens}{2007}]{spekkens_evidence_2007}
Spekkens, R.W. 2007, March.
\newblock Evidence for the epistemic view of quantum states: {A} toy theory.
\newblock {\em Physical Review A\/}~{\em 75\/}(3): 032110.
\newblock \doi{10.1103/PhysRevA.75.032110} .

\bibitem[\protect\citeauthoryear{Stalnaker}{Stalnaker}{2006}]{stalnaker_logics_2006}
Stalnaker, R. 2006.
\newblock On {Logics} of {Knowledge} and {Belief}.
\newblock {\em Philosophical Studies: An International Journal for Philosophy
  in the Analytic Tradition\/}~{\em 128\/}(1): 169--199 .

\bibitem[\protect\citeauthoryear{Sudbery}{Sudbery}{2019}]{sudbery_hidden_2019}
Sudbery, A. 2019.
\newblock The hidden assumptions of {Frauchiger} and {Renner}.
\newblock {\em International Journal of Quantum Foundations\/}~5: 98--109 .

\bibitem[\protect\citeauthoryear{Tennant}{Tennant}{1997}]{tennant_taming_1997}
Tennant, N. 1997.
\newblock {\em The {Taming} of the {True}}.
\newblock Oxford University Press.

\bibitem[\protect\citeauthoryear{{van Ditmarsch}, Halpern, {van der Hoek}, and
  Kooi}{{van Ditmarsch} et~al.}{2015}]{van_ditmarsch_handbook_2015}
{van Ditmarsch}, H., J.~Halpern, W.~{van der Hoek}, and B.~Kooi. 2015.
\newblock {\em Handbook of {Epistemic} {Logic}}.
\newblock College Publications.

\bibitem[\protect\citeauthoryear{{van Ditmarsch}, {van Der Hoek}, and
  Kooi}{{van Ditmarsch} et~al.}{2007}]{van_ditmarsch_dynamic_2007}
{van Ditmarsch}, H., W.~{van Der Hoek}, and B.~Kooi. 2007.
\newblock {\em Dynamic epistemic logic}, Volume 337.
\newblock Springer Science \& Business Media.

\bibitem[\protect\citeauthoryear{Vilasini, Nurgalieva, and {del Rio}}{Vilasini
  et~al.}{2019}]{vilasini_multi-agent_2019}
Vilasini, V., N.~Nurgalieva, and L.~{del Rio}. 2019.
\newblock Multi-agent paradoxes beyond quantum theory.
\newblock {\em New Journal of Physics\/}~{\em 21\/}(11): 113028.
\newblock \doi{10.1088/1367-2630/ab4fc4} .

\bibitem[\protect\citeauthoryear{Von~Neumann}{Von~Neumann}{1955}]{von_neumann_mathematical_1955}
Von~Neumann, J. 1955.
\newblock {\em Mathematical foundations of quantum mechanics}.
\newblock Number~2. Princeton university press.

\bibitem[\protect\citeauthoryear{Waaijer and {van Neerven}}{Waaijer and {van
  Neerven}}{2021}]{waaijer_relational_2021}
Waaijer, M. and J.~{van Neerven}. 2021.
\newblock Relational analysis of the {Frauchiger}--{Renner} paradox and
  interaction-free detection of records from the past.
\newblock {\em Foundations of Physics\/}~{\em 51\/}(45).
\newblock \doi{https://doi.org/10.1007/s10701-021-00413-4} .

\bibitem[\protect\citeauthoryear{Wallace}{Wallace}{2012}]{wallace_emergent_2012}
Wallace, D. 2012.
\newblock {\em The {Emergent} {Multiverse}: {Quantum} {Theory} according to the
  {Everett} {Interpretation}}.
\newblock Oxford: Oxford University Press.

\bibitem[\protect\citeauthoryear{Williamson}{Williamson}{1982}]{williamson_intutionism_1982}
Williamson, T. 1982.
\newblock Intutionism {Disproved}?
\newblock {\em Analysis\/}~{\em 42\/}(4): 203--207.
\newblock \doi{10.2307/3327773} .

\bibitem[\protect\citeauthoryear{Williamson}{Williamson}{1987}]{williamson_paradox_1987}
Williamson, T. 1987.
\newblock On the {Paradox} of {Knowability}.
\newblock {\em Mind\/}~{\em 94\/}(382): 256--261 .

\end{thebibliography}


%\newpage
%\printbibliography

\end{document}